\documentclass[]{article}
\usepackage{amssymb}
\begin{document}
\title{A geometric procedure for the reduced-state-space quantisation of constrained systems}
\author{Charis Anastopoulos \thanks{anastop@physics.upatras.gr}, \\
Department of Physics, University of Patras, \\
26500 Patras, Greece } \maketitle

\renewcommand {\thesection}{\arabic{section}}
 \renewcommand {\theequation}{\thesection. \arabic{equation}}
\let \ssection = \section
\renewcommand{\section}{\setcounter{equation}{0} \ssection}

\begin{abstract}
We propose in this paper a new method for the quantisation of
systems with first-class constraints. This method is a combination
of the coherent-state-path-integral quantisation developed by
Klauder, with the ideas of reduced state space quantisation. The
key idea is that the physical Hilbert space may be defined by a
coherent-state  path-integral on the reduced state space and that
the metric on the reduced state space that is necessary for the
regularisation of the path-integral may be computed from the
geometry of the classical reduction procedure. We provide a number
of examples--notably the relativistic particle. Finally we discuss
the quantisation of systems, whose reduced state space has
orbifold-like singularities.

\end{abstract}

\section{Introduction}

 There exist many approaches for the quantisation of systems
with (first-class) constraints. They fall essentially into two
categories, according to whether they implement the constraints
before or after quantisation. In the first category lie schemes
that attempt to quantise directly the reduced state space of a
constrained system. Schemes of the second category are exemplified
by Dirac quantisation:  one quantises the theory before the
imposition of the constraints and  identifies the Hilbert space
for the physical degrees of freedom as the zero-eigenspace of a
quantum operator that represents the constraint.

The present paper proposes a variation of the Reduced State Space
Quantisation scheme. It is based upon coherent-state-path-integral
quantisation scheme method, which has been developed by Klauder
\cite{KlDa84, Kla88}. In this method the only input necessary for
the quantisation of a classical symplectic manifold $\Gamma$ is
the introduction of a Riemannian metric $ds^2$ on $\Gamma$. The
metric allows the definition of a Wiener process on $\Gamma$,
which plays the role of a regulator for the rigorous definition of
a path-integral for the coherent state matrix elements. The key
suggestion of our proposal is that a Riemannian metric on the
reduced state space may be always identified, by exploiting
 the {\em
geometry} associated to the classical reduction procedure.

\section{Dirac vs Reduced state space quantisation} A Hamiltonian
system with first-class constraints is described by a symplectic
manifold $\Gamma$, equipped with a symplectic form $\Omega$ and a
number of constraint functions $f^i : \Gamma \rightarrow {\bf R}$,
such that $\{f^i, f^j \} = c^{ij}{}{}_k f^k$, for some structure
{\em functions} $c^{ij}{}{}_k$. One defines the constraint surface
$C$ as the submanifold of $\Gamma$ defined by the conditions $f^i
= 0$. The restriction of the symplectic form $\Omega$ on $C$ is
degenerate. The reduced state space $\Gamma_{red}$ is defined as
the space of all orbits on the constraint surface generated by the
constraints through the Poisson bracket. In other words,
$\Gamma_{red}$ is the quotient $C/ \sim$, where $\sim$ is an
equivalence relation, defined as follows: $z \sim z'$, if $z'$ can
be reached from $z$ through a canonical transformation generated
by the constraints.

We define the projection map $\pi:C \rightarrow \Gamma_{red}$ as
$\pi(z) = [z]$, where $[z]$ the orbit, in which the point $z \in
C$ belongs then. $\Gamma_{red}$ is equipped with a symplectic form
$\tilde{\Omega}$, such that $\pi_*\tilde{\Omega} = \Omega|_C$.

The physical idea reduced state space quantisation (RSSQ) is that
the rules of quantisation are only meaningful, when applied to the
true physical degrees of freedom and are rather ambiguous when
applied to the   gauge degrees of freedom. It is suggested
therefore that one should  construct the quantum theory for the
reduced state space $\Gamma_{red}$.  In general, $\Gamma_{red}$
has a non-trivial topological structure, and generalisations of
the canonical quantisation scheme is needed to pass into the
quantum theory. The implementation of RSSQ, however, necessitates
that one is able to fully solve the constraints classically,
something that is very difficult if not impossible for interesting
physical systems (e.g. general relativity). So for such systems
not even the first step in the quantisation procedure can be
implemented, leaving us  in a complete blank about the properties
of various quantum mechanical kinematical variables.

Moreover, the reduced state space of field systems characterised
by first-class constraints generically consists of non-local
variables: the true degrees of freedom are not fields themselves.
For this reason the spacetime character of the theory is not
explicitly manifest in the reduced state space. In a reduced state
space quantisation the quantum mechanical variables one constructs
are not fields. Furthermore, there exist quantities of potential
interest quantum mechanically that do not commute with the
constraints--such as the area and volume operators  in loop
quantum gravity \cite{RoSm95}. There is no way to describe such
objects in the RSSQ scheme.

 In many
systems of interest, such as General Relativity, the reduced state
space is not a manifold but an orbifold:  a singular structure
arises as a result of taking the equivalence classes with respect
to the action of the constraints. This creates additional
technical problems in the quantisation of such systems.

Dirac quantisation on the other hand involves the construction a
{\em kinematical} Hilbert space $H$, in which the basic
observables of $\Gamma$ are represented by self-adjoint operators.
One then constructs self-adjoint operators $\hat{f}^i$ to
represent the constraints. The physical Hilbert space $H_{phys}$
is the subspace of $H$ consisting of all vectors $| \psi \rangle
\in H$, such that $\hat{f}^i | \psi \rangle = 0$. Typically
variations of this theme are employed for the construction of the
physical Hilbert space, because constraint operators do not have
continuous spectrum around zero. Hence `wave functions' that solve
the quantum constraints are not square-integrable. This issue
together with the normal-ordering ambiguities associated with the
definition of constraint operators are the major technical
problems of Dirac quantisation.

The method we propose here solves the constraints at the classical
level, but it does allow for the existence of a kinematical
Hilbert space (like Dirac quantisation) and of a map between that
Hilbert space and the one corresponding to the physical degrees of
freedom.

\section{Geometric reduction and coherent state path integrals}
\subsection{Coherent states and path-integrals}
Our prescription for RSSQ is based upon Klauder's scheme of
coherent state quantisation. The Hilbert space $H$ associated to a
classical symplectic manifold $\Gamma$ with symplectic form
$\Omega$ is a suitable subspace of the space ${\cal L}^2(\Gamma,
dz)$, where $dz$ is a shorthand for the Liouville form $ \Omega
\times \ldots \times \Omega$ on $\Gamma$. This subspace
corresponds to a projection operator ${\bf E}$ on ${\cal
L}^2(\Gamma, dz)$, whose matrix elements define a complex-valued
function $K(z,z')$, $z, z' \in \Gamma$, such that
\begin{eqnarray}
\int dz' K(z,z') K(z',z'') = K(z,z''), \label{1}\\
K(z,z') = K^*(z',z). \label{2}
\end{eqnarray}
Moreover, the Hilbert space $H$ is equipped with a family of
generalised coherent states $| z \rangle, z \in \Gamma$, such that
\begin{eqnarray}
\langle z| z' \rangle = K(z,z'). \label{3}
\end{eqnarray}

Conversely, given any complex-valued function on $\Gamma \times
\Gamma$ that satisfies (\ref{2}) and is positive definite, one may
construct uniquely a Hilbert space $H$, equipped with a family of
generalised coherent states $|z \rangle$, such that (\ref{3})
holds--see \cite{Kla97} for a detailed exposition. Positivity in
this context is defined as the requirement that
\begin{eqnarray}
\sum_{l,l'} c_l c^*_{l'} K(z_l,z_{l'}), \geq 0, \label{positivity}
\end{eqnarray}
 for any finite set of complex number $c_l$ and state space points
 $z_l$, and where equality in (\ref{positivity}) holds only for $c_l = 0$

It is important to note that two kernels $K_1$ and $K_2$ related
by a phase transformation
\begin{eqnarray}
K_1(z,z') = e^{i \theta(z) - i \theta(z')} K_2(z,z'),
\end{eqnarray}
define the same Hilbert space, since the corresponding family of
coherent states differ only with respect to the phase $e^{i
\theta}$.

 The overlap kernel may be computed by means of
path-integrals, in a procedure developed by Klauder. The key
ingredient is a Riemannian metric $ds^2$ on the classical phase
space. This metric is employed for the purpose of regularisation
of the path-integral expressions, which cannot otherwise be
defined.

The metric $ds^2$ defines a Wiener measure on the space of paths
$\Pi$, which consists of suitable continuous functions (in general
non-differentiable) from a subset of ${\bf R}$ to $\Gamma$.
\begin{eqnarray}
d \mu^{\nu}[z(\cdot)] = Dz^a (\cdot) e^{- \frac{1}{2 \nu} \int_0^T
d \lambda \left(\frac{ds}{d\lambda} \right)^2 d \lambda},
\end{eqnarray}
where $\nu$ is a diffusion coefficient for the Wiener process,
which plays the role of a regularisation parameter.

 The overlap kernel is then obtained as
 \begin{eqnarray}
\langle z_f|z_i \rangle = \lim_{\nu \rightarrow \infty} {\cal
N}_{\nu} \int d \mu^{\nu}[z(\cdot)]e^{i \int A[z(\cdot)]}.
\label{pathintegral}
 \end{eqnarray}
 In this expression the summation is over paths $z(\cdot)$, such
 that $z(0) = z_i$ and $z(T) = z_f$. The Wiener measure is
 conditioned on these boundary values. The object $A_a$ is a U(1)
 connection, which is related to the symplectic form of the
 manifold $\Gamma$ by means of the relation $\Omega = dA$. In the
 coordinates that $\Omega = dp_a \wedge dq^a$, the connection $A$
 reads $A = p_a dq^a$. The parameter ${\cal N}_{\nu}$ is a
 normalisation constant, which may be fixed by the condition that
 $\langle z|z \rangle = 1$.

The propagator has also a path integral expression
 \begin{eqnarray}
\langle z_f|e^{-i \hat{H}T}|z_i \rangle = \lim_{\nu \rightarrow
\infty} {\cal N}_{\nu}\int d \mu^{\nu}[z(\cdot)]e^{i \int d A_a(z)
dz^a - i \int_0^T d \lambda h(z) },
 \end{eqnarray}
where $h(z)$ is the classical Hamiltonian function corresponding
to the Hamiltonian operator $\hat{H}$.

Klauder  extended this procedure to deal with systems with
constraints (either first-class or second-class) \cite{Kla97,
Klaother}. His scheme involved a generalisation of the Dirac
method-- he showed that the projection operator into the zero
eigenspace of the constraint operators may be expressed as a
path-integral over the space of paths on $\Gamma$.

In this paper we suggest that it is not necessary to define the
constraints at the quantum level. Rather the reduction may be
implemented at the classical level. in a way that allows the
definition of  integral of the form (\ref{pathintegral}), with
paths on the {\em reduced state space}.

\subsection{Path-integrals on the reduced state space}

Our proposal starts from the remark that the path-integral
(\ref{pathintegral}) may be employed to also define coherent
states on the reduced state space $\Gamma_{red}$. The symplectic
potential $A$ on the reduced state space is easily obtained, since
the reduced state space $\Gamma_{red}$ is also a symplectic
manifold. The problem is to identify the correct metric
$ds^2_{red}$ on $\Gamma_{red}$, through which to define the
path-integral.

In unconstrained systems the metric is chosen by  symmetry
requirements. It is usually a homogeneous metric, whose isometries
corresponding to the transitive action of a  group on the state
space. For classical mechanics on ${\bf R}^{2n}$, the
corresponding group is the Weyl group, for the sphere $S^2$ that
corresponds to a spin-system it is the group $SU(2)$ etc. This
group--referred to as the canonical group-- provides a symmetry of
the kinematical description \cite{Ish84}. Conversely, the
requirement of symmetry in the kinematical description allows one
to select a homogeneous Riemannian metric to define the
path-integral.

But a symmetry of the kinematical description does not, in
general, leave the dynamics or the constraints invariant. (We know
in standard quantum mechanics that no physically interesting
Hamiltonian commutes with both position and momentum that form the
generators of the Weyl group.) The metric $ds^2_{red}$ on
$\Gamma_{red}$ therefore cannot be easily identified from the
`symmetries' of the original theory. Still $ds^2_{red}$ may be
computed, if one has chosen a Riemannian metric $ds^2$ on the
unconstrained state space $\Gamma$, which can itself be chosen by
invoking symmetry arguments.

A metric $ds^2$ on $\Gamma$ defines by restriction a metric on the
constraint surface $C$--hence a distance function $d(z,z')$
between points $z, z' \in C$. By definition, a point $\zeta$ of
$\Gamma_{red}$ is a collection of points on $C$ generated by the
action  of the constraint functions. Hence a point $\zeta$ is
identified with a submanifold of $C$. Moreover the submanifolds
corresponding to different points of the reduced state space are
disjoint, as different equivalence classes are always disjoint.
This implies that one may exploit the distance function on $C$ to
define a distance function on $\Gamma_{red}$, namely
\begin{eqnarray}
d_{red}(\zeta, \zeta') = \inf_{z \in \zeta, z' \in \zeta'}
d(z,z'). \label{distance}
\end{eqnarray}
From $d_{red}$ one may define a metric on $\Gamma_{red}$.

To summarise, given a metric on $\Gamma$--which defines a
quantisation of the system prior to the imposition of the
constraints-- one may construct uniquely a metric on
$\Gamma_{red}$, through which one may construct the Hilbert space
for the true degrees of freedom.

There is a point of ambiguity in the present construction. The
definition of distance in the reduced state space Eq.
(\ref{distance}) involves the infimum of the distances between the
points of these orbits. The infimum involves taking a limit, and
it may turn out that the distance
 between specific orbits $\zeta$
and $\zeta'$ vanishes--even though these orbits are disjoint. In
that case one has to identify the points $\zeta$ and $\zeta'$ in
the reduced state space--the coherent states cannot distinguish
between them. In many cases this point turns out to be a benefit,
rather than a disadvantage. We shall later show that it is usually
in singular points (ones that render $\Gamma_{red}$ into an
orbifold) that the distance function vanishes.

There is a physical interpretation of the reduction procedure we
propose, which has to do with the uncertainty principle. A
Riemannian metric on state space may be employed for a geometric
expression of the uncertainty principle \cite{AnSav03}. Two points
on the classical state space may be distinguished quantum
mechanically only if their distance $\delta s^2$ is greater than
one (in units such that $\hbar = 1$). The presence of a
first-class constraint implies that points in the same orbit
cannot be physically distinguished. To see whether one can
distinguish between two orbits, the distance between any two
points of each orbit should be larger than one (in terms of the
original metric). This suggests the definition of  the statistical
distance between orbits--and consequently the metric on
$\Gamma_{red}$-- as the minimum of distance between its points.

\subsection{Relation to Dirac quantisation}

An important feature of the construction described above is that
it  preserves some of the advantages of Dirac quantisation, namely
the existence of quantum operators for variables that do not
commute with the constraints. Indeed, one may employ the metric
$ds^2$ on $\Gamma$ to construct a kinematical Hilbert space $H$,
spanned by a coherent states family $|z \rangle$, $z \in \Gamma$.
We may consider then the subspace of $H$, spanned by all finite
linear combinations $\sum_{l = 1}^{n} c_l |z_l \rangle$, where the
poitns $z_l \in C$. On the other hand, the physical Hilbert space
$H_{phys}$ is constructed from the metric $ds^2_{red}$ and is
spanned by coherent states $|\zeta \rangle$. The natural
projection map $\pi: C \rightarrow \Gamma_{red}$ defined as
$\pi(z) = [z]$, induces a  map $i_{\pi}$ between $H_C$ and
$H_{phys}$
\begin{eqnarray}
i_{\pi} |z \rangle = | [z] \rangle, \label{map}
\end{eqnarray}
which by linearity can be extended to the whole of $H_C$.
 Moreover, if we denote by $E$ the
projection map from $H$ to $H_C$, any vector $| \psi \rangle$ on
$H$ may be mapped to a vector on $H_{phys}$, by mapping its
projection $P | \psi \rangle$ on $H_C$ to $H_{phys}$ through the
map $i_{\pi}$. The map $i_{\pi}$ is then extended to a map between
the kinematical Hilbert space and the physical Hilbert space. It
is  not a projection operator in general (or even a self-adjoint
operator), as the inner products on $H$ and on $H_C$ cannot be
simply related \footnote{Note that even in conventional Dirac
quantisation the mapping from $H$ and $H_{phys}$ is not
implemented by a self-adjoint operator, if the constraints do not
have a continuous spectrum at 0. }. Still $i_{\pi}$ may be
employed to distinguish the 'gauge invariant part' of any operator
in $H$, namely the part of the operator that can be 'projected' to
the physical Hilbert space. Thus we obtain some of the benefits of
Dirac quantisation, without having to deal with issues such as
operator ordering or the existence of a continuous spectrum for
the constraints near zero.

The map between $H$ and $H_{phys}$ is hardly unique--the physical
predictions of the theory remain the same, if one multiplies the
right-hand-side of (\ref{map}) with a $z$-dependent phase $e^{i
\theta(z)}$. Each choice of a function $e^{i \theta(z)}$ leads to
a different linear map. But this arbitrariness is not problematic,
because in any case the definition of $H$ in most physical systems
contains a large degree of arbitrariness--especially in field
theories.

In a nutshell, our procedure proposes a way to pass from the
symplectic form $\Omega$ and Riemannian metric $ds^2$ on $\Gamma$
to the symplectic form $\Omega_{red}$ and Riemannian metric
$ds^2_{red}$ on $\Gamma_{red}$. The knowledge of the symplectic
form and the metric defines uniquely a Hilbert space equipped with
a family of coherent states. We may then construct on such Hilbert
space $H$ for $\Gamma$ constructed via path integrals from
$\Omega$ and $ds^2$ and one Hilbert space $H_{phys}$ for
$\Gamma_{red}$ constructed from $\Omega_{red}$ and $ds^2_{red}$.
The reduction procedure that takes us from $(\Omega, ds^2)$ to
$(\Omega_{red}, ds^2_{red})$ defines a map from $H$ to $H_{phys}$,
in a way that mirrors the projection map from $H$ to $H_{phys}$
appearing in Dirac quantisation.

\subsection{Relation with path-integrals on $\Gamma$}

An important point that provides a clarification of our proposal
is that the  path integral over the reduced state space may be
equivalently  rewritten as one over paths on $\Gamma$, similar
(but not identical) to the one employed by Klauder in the Dirac
quantisation scheme through the coherent-state-path-integral.

If we define the Wiener measure $d \tilde{\mu}[\zeta(\cdot)]$ over
paths in $\Gamma_{red}$ through the metric $ds^2_{red}$ and denote
as $\tilde{A}$ a U(1) connection on $\Gamma_{red}$ corresponding
to $\Omega_{red}$, one writes the path-integral
(\ref{pathintegral}) on the reduced state space as
\begin{eqnarray}
 \langle \zeta_f|\zeta_i
\rangle = \lim_{\nu \rightarrow \infty} {\cal N}_{\nu} \int d
\tilde{\bar{\mu}}^{\nu}[\zeta(\cdot)]e^{i \int
\tilde{A}[\zeta(\cdot)]}. \label{pathintegral2}
\end{eqnarray}

Assuming that $\Gamma$ is a 2n-dimensional manifold, and that we
have $m < n$ first-class constraints, the constraint surface $C$
is a $(2n - m)$ dimensional manifold, and $\Gamma_{red}$ has
dimension equal to $2(n-m)$. If the functions $\zeta^i, i = 1,
\ldots, 2(n-m)$ define a coordinate system on $\Gamma_{red}$,
their pull-back on $C$ define a coordinate
 system on $C$
together with $m$ coordinates $v^i, i= 1, \ldots, m$ that span
each gauge orbit. The symplectic form on $C$ does not depend on
the coordinates $\lambda^i$--it is  written as
\begin{eqnarray}
\Omega|_C = \Omega_{ij}(\zeta) d \zeta^i \wedge d \zeta^j
\end{eqnarray}

If $\tilde{A}^i$ is a $U(1)$ connection on $\Gamma_{red}$ that
satisfies $d\tilde{A} = \Omega$, the most general connection
one-form on $C$ that satisfies $dA = \Omega|_C$ may be written
locally as
\begin{eqnarray}
\tilde{A}_i(\zeta) d \zeta^i + d \theta(\zeta,v),
\end{eqnarray}
in terms of a scalar function $\theta$ on $C$.

Let us now consider an integral over paths on $C$
\begin{eqnarray}
\int d \mu_{C}^{\nu}[z(\cdot)]  e^{i \int A[z(\cdot)]}
\end{eqnarray}
Let us, for the moment, refrain from specifying the metric on $C$
that determines the Wiener measure $d \mu_{\nu}[z(\cdot)]$, except
for the fact that it is conditioned on the endpoints. The exponent
in the path integral reads $ i \int \tilde{A}_i d \zeta^i + i
(\theta(\zeta_f,v_f) - \theta(\zeta_i,v_i)$. The path integral
then reads
\begin{eqnarray}
e^{i\theta(\zeta_f,v_f) - i \theta(\zeta_i,v_i)} \int d
\mu^{\nu}[z(\cdot)]  e^{i \int A_i d \zeta^i},
\end{eqnarray}
since the integration measure is conditioned at the endpoints. The
phases in front of the path-integral may be reabsorbed in a phase
change of the coherent states, and may therefore be dropped out.

If we assume that the integration measure factorises into a piece
along the orbits ($\mu_{gauge}$) and one across the orbits
$\mu_{red}$)
\begin{eqnarray}
\int d \mu_{C}^{\nu}[z(\cdot)] = \int d \mu_{red}^{
\nu}[\zeta(\cdot)] \int d \mu_{gauge}^{\nu}[v(\cdot)]
\label{factor}
\end{eqnarray}
then we see that
\begin{eqnarray}
\int d \mu_{C\nu}[z(\cdot)]  e^{i \int A_a dz^a} = \int d
\mu_{red}^{\nu}[\zeta(\cdot)] e^{i \int A_i d \zeta^i} \int d
\mu_{gauge}^{\nu}[v(\cdot)]
\end{eqnarray}
If the two measures are separately normalised,  the overlap kernel
on the reduced phase space may be obtained by a path integral on
the constraint surface,
\begin{eqnarray}
\langle \zeta_f|\zeta_i \rangle = \lim_{\nu \rightarrow \infty}
{\cal N}_{\nu} \int d \tilde{\mu}^{\nu}_{ C}[z(\cdot)] e^{i \int
A[z(\cdot)]}.
\end{eqnarray}
The crucial point is that the measure should satisfy the
factorisation condition (\ref{factor}), which involves a suitable
choice of metric on $C$. Recall that the metric is a bilinear
functional on the tangent space $T_zC$ of the constraint surface.
The degeneracy of the symplectic form $\Omega$ implies that each
tangent space is split into  the degeneracy subspace $D_zC$ of all
tangent vectors $X$ such that $\Omega_C(X,\cdot) = 0$ and its
complement $\bar{D}_zC$, which may be naturally identified with
$T_{\zeta}\Gamma_{red}$. Since the Wiener measure is of the form
(\ref{factor}) the factorisation condition may be obtained if the
metric is in a block-diagonal form with respect to the split $T_zC
= D_zC \oplus \bar{D}_zC$. In other words, g(X,Y) = 0 , if $X \in
D_zC$ and $Y \in \bar{D}_zC$. Moreover, it is necessary that the
restriction of the metric in $\bar{D}_zC$ does not depend on the
variables $v$. The Hamiltonian vector fields generated by the
constraints should leave the restriction of the metric on
$\bar{D}_zC$ invariant.

The constraint surface $C$ is defined as the submanifold of
$\Gamma$, in which the constraint functions $f^a, a = 1, \ldots,
m$ vanish. Any ordinary integral over $C$, may be expressed as an
integral over $\Gamma$ by the inclusion of the product of delta
functions $\prod_a \delta(f^a(z))$. Similarly, we may turn the
path integral on $C$ into a path integral over $\Gamma$ by the
insertion of delta function at each time point of the
discretisation.

We then write
\begin{eqnarray}
\langle \zeta_f|\zeta_i \rangle = \lim_{\nu \rightarrow \infty}
{\cal N}_{\nu} \int d \mu^{\nu }[z(\cdot)] \prod_a \Delta[f^a]
e^{i \int A[z(\cdot)]},
\end{eqnarray}
where $\Delta[f^a] = \prod_t \delta (f^a(z(t)))$; the product
refers to any discretisation of the paths in the path integral.
The integration measure may be defined by any metric $ds^2$ on
$\Gamma$, which reduces to a factorisable metric on $C$.
Similarly, the connection on $\Gamma$ may be chosen arbitrarily as
long as it satisfies $dA = \Omega$.

Exploiting the representation of the delta function as an integral
$\delta(f) = \int d N e^{-iNF}$ we  write formally
\begin{eqnarray}
\Delta[f^a] = \int DN(\cdot) e^{-i dt N_a(t) f^a(t)},
\end{eqnarray}
where $DN(\cdot)$ is a continuous limit of $\prod_a \prod_t
dN^a_t$.
 We then write the formal expression
\begin{eqnarray}
\langle \zeta_f|\zeta_i \rangle = \lim_{\nu \rightarrow \infty}
{\cal N}_{\nu} \int DN(\cdot) \int d \mu_{\nu }[z(\cdot)] e^{i
\int A_a dz^a - i \int_0^T ds N_a(s) f^a(s)}
\end{eqnarray}
This expression is {\em formally similar} to the one employed by
Klauder in his implementatation of Dirac quantisation through the
coherent state path-integral \cite{Kla97}. The key difference is
that the Wiener process in Klauder's scheme  is defined by means
of a homogeneous metric on $\Gamma$, while in the RPSQ scheme the
metric on $\Gamma$ is in general non-homogeneous and needs to
satisfy the factorisation condition on the constraint surface we
described earlier\footnote{Moreover, in Klauder's version of Dirac
quantisation the measure over $N(\cdot)$ is normalised to one,
while here it is only a formal continuum limit of $\prod_t dN_t$}.
The difference in the metrics  implies that the  diffusion
processes regularising the path-integral are different, hence the
final expressions for the overlap kernels. Since, however, the
metric only appears for regularisation purposes, and both the
constraints and the symplectic form are the same, the two methods
are expected to yield the same results at the semi-classical
level.

\section{Examples}

\subsection{A trivial example}
We may consider a particle moving on the Euclidean three-space
${\bf R}^3$. The state space $\Gamma = {\bf R}^6$  is spanned by
the coordinates $(x^i,p_i), i = 1,2,3$ and is equipped with the
natural symplectic form. We then consider the constraint $x^3=0$,
which can be trivially shown to imply that $\Gamma_{red} = {\bf
R}^4$, spanned by the variables $(x_1,x_2,p_1,p_2)$.

The coherent states on $\Gamma$ are the standard Gaussian coherent
states, with overlap kernel
\begin{eqnarray}
\langle {\bf x}, {\bf p}| {\bf x'}, {\bf p'} \rangle = \exp \left(
ip_ix'^i - i p_i' x^i -\frac{1}{2} |{\bf x} - {\bf x'}|^2 -
\frac{1}{2}|{\bf p} - {\bf p'}|^2 \right)
\end{eqnarray}

which correspond to the homogeneous Riemannian metric on ${\bf
R}^6$:
\begin{eqnarray}
ds^2_{\Gamma} = \delta_{ij} dx^i dx^j + \delta^{ij} dp_i dp_j
\end{eqnarray}
Since the orbits of the constraint surface are the lines of
constant $(x,y,p_x,p_y)$ with $z = 0$, it is elementary to show
that the distance between the lines corresponds to the reduced
metric
\begin{eqnarray}
ds^2_{\Gamma_{red}} = dx_1^2 + dx_2^2  + dp_1^2 + dp_2^2,
\end{eqnarray}
which again corresponds to Gaussian coherent states for the
reduced Hilbert space.

\subsection{A spin system}
We consider the state space ${\bf R}^4$, with variables
$(x_1,x_2,p_1,p_2)$, equipped with the standard symplectic form
\begin{eqnarray}
\omega = dp_1 \wedge dx_1 + dp_2 \wedge dx_2
\end{eqnarray}
and a constraint
\begin{eqnarray}
\frac{1}{2} (p_1^2 +p_2^2 + x_1^2 +x^2_2) = k > 0.
\end{eqnarray}
The constraint surface is the two sphere $S^3$. Employing the
coordinates $(\theta, \phi, \chi)$ on $S^3$ through the definition
\begin{eqnarray}
\frac{1}{\sqrt{2}} (x_1 - i p_1) = \sqrt{k} \cos \frac{\theta}{2}
e^{i(\phi + \chi)/2} \\
\frac{1}{\sqrt{2}} (x_2 - i p_2) = \sqrt{k} \sin \frac{\theta}{2}
e^{i(\phi - \chi)/2},
\end{eqnarray}
we obtain
 \begin{eqnarray}
\Omega_C = \frac{k}{2} \sin \theta d \theta \wedge d \phi,
 \end{eqnarray}
which implies that the degenerate direction corresponds to the
vector field $\frac{\partial}{\partial \chi}$. Its orbits define
the usual Hopf fibration of $S^3$, hence the reduced state space
is $S^2$ equipped with the standard symplectic form. It is the
state space of a classical spin system. As is well-known, the
single-valuedness of the $U(1)$ connection $\cos \theta d \phi$
that appears in the path-integral, implies that in the quantum
theory $k = 2 n$, for integers $n$.

The  homogeneous metric on ${\bf R}^4$ corresponding to Gaussian
coherent states is
\begin{eqnarray}
ds^2  = dx_1^2 + dx^2_2 + dp_1^2 + dp_2^2,
\end{eqnarray}
reduces on the constraint surface $S^3$ to
\begin{eqnarray}
ds^2_C = \frac{k}{2}(d \theta^2 + d \phi^2 + d \chi^2 - 2 \cos
\theta d \phi d \chi).
\end{eqnarray}
It is easy to minimise this metric over the orbits of constant
$(\theta, \phi)$ to obtain the Riemannian metric on ${\bf S^2}$.
\begin{eqnarray}
ds^2_{red} = \frac{k}{2} (d \theta^2 + \sin^2 \theta d \phi^2).
\end{eqnarray}
This is the {\em homogeneous} metric for a sphere of radius $k/2 =
n$; for this metric and this connection one may obtain through the
path-integral (\ref{pathintegral}) the quantum description of spin
in terms of the irreducible representations of $SU(2)$ for each
value of $n$ \cite{Kl81}.

\subsection{Spinless relativistic particle}

The relativistic particle illustrate the strengths of the method,
perhaps better than any other example. The reason is that the
Poincar\'e symmetry of the system leaves few alternatives about
the form of the Riemannian metric on the constraint surface, The
resulting coherent states are therefore fully covariant.

The spinless relativistic particle of mass is described by the
presymplectic manifold $C$ (the constraint surface), which is
spanned by a {\em unit, future-directed, timelike vector} $I$ and
an element $X$ of Minkowski spacetime. The topology of $C$ is then
$V \times {\bf R}^4$, where $V$ the mass-shell for particles of
mass $m$. The symplectic form then reads
\begin{eqnarray}
\Omega_C = - m dI_{\mu} \wedge dX^{\mu}.
\end{eqnarray}
Note that we employ a $(+---)$ signature convention. It is easy to
see that the vector field $I^{\mu} \frac{\partial}{\partial
X^{\mu}}$ corresponds to the null direction of $\Omega_C$--it is
the Hamiltonian vector field generated by the constraint $I^2 - 1
= 0$. The parameter $u = I\cdot X$ is the gauge degree of freedom
along the orbits of the constraint, while the variables $I_{\mu},
Y^{\mu} = X^{\mu} - u I^{\mu}$ are projected on the reduced state
space $\Gamma_{red}$. The reduced symplectic form reads
\begin{eqnarray}
\Omega_{red} = - m dI_{\mu} \wedge dY^{\mu}.
\end{eqnarray}
We next identify a Lorentz-invariant Riemannian metric on $C$. For
this purpose we need to identify a Lorentz-invariant notion of
distance between pairs $(I, X)$ and $(I', X')$. On the mass-shell
$V$ there exists a natural Riemannian metric
\begin{eqnarray}
ds^2_V = - dI^{\mu} dI_{\mu} \geq 0,
\end{eqnarray}
which may be employed to define  distance between $I$ and $I'$. To
define a (positive definite) distance between $X$ and $X'$, let us
note that if $I = I'$, then $2I_{\mu} I_{\nu} + \eta_{\mu \nu}$
defines a Riemannian metric on ${\bf R}^4$, and the corresponding
distance between $X$ and $X'$ equals
\begin{eqnarray}
[2I_{\mu} I_{\nu} - \eta_{\mu \nu}](X^{\mu} - X'^{\mu}) (X^{\nu} -
X'^{\nu}). \label{dist}
\end{eqnarray}
If $I \neq I'$, we need to boost $X'$ to the frame where $I = I'$
and then employ expression (\ref{dist}) for distance. If we denote
by $\Lambda$ the unique Lorentz transformation (boost) that takes
$I$ into $I'$, we may define the distance function
\begin{eqnarray}
[2I_{\mu} I_{\nu} - \eta_{\mu \nu}](X^{\mu} -
\Lambda^{\mu}{}_{\rho} X'^{\rho}) (X^{\nu} -
\Lambda^{\nu}{}_{\sigma} X'^{\sigma}).
 \end{eqnarray}
The construction above suggests the following definition of a
Lorentz invariant metric on $C$
\begin{eqnarray}
ds^2_C = - c_1 dI^{\mu} dI_{\mu}   + \frac{c_2}{m}[2I_{\mu}
I_{\nu} - \eta_{\mu \nu}] (dX^{\mu} -I^{\mu} X\cdot dI - dI^{\mu}
I \cdot X)\nonumber \\ \times (dX^{\nu} -I^{\nu} X\cdot dI -
dI^{\nu} I \cdot X), \label{metric}
\end{eqnarray}
where $c_1, c_2$ are arbitrary positive numbers and the mass $m$
appears in the denominator to make the metric unit dimensionless
($\hbar = c = 1$). We shall choose for convenience $c_1 = c_2 =
\frac{1}{2}$.

To construct a metric on $\Pi_{red}$, we write Eq. (\ref{metric})
in terms of the  coordinate $u$, together with the physical
variables $I^{\mu}, Y^{\mu}$
\begin{eqnarray}
ds^2_C =  - \frac{1}{2} dI^{\mu} dI_{\mu}  + \frac{m^2}{2}[I_{\mu}
I_{\nu} - \eta_{\mu \nu}] dY^{\mu} dY^{\nu} + \frac{m^2}{2} du^2
\end{eqnarray}
It is then easy to find the distance between neighboring orbits,
by minimising over $du$, thus obtaining a Lorentz-invariant metric
on the reduced state space
\begin{eqnarray}
ds^2_{red} =  - \frac{1}{2} dI^{\mu} dI_{\mu}  + m^2 [I_{\mu}
I_{\nu} - \eta_{\mu \nu}] dY^{\mu} dY^{\nu} \label{metred}.
\end{eqnarray}
\paragraph{Particle in 2d} For simplicity we next consider the case of a particle in  two-dimensional
Minkowski spacetime. In terms of $I = I^1$ and $q = I^0 X^1 - I^1
X^0$, we may write $\Omega_{red}$ as
\begin{eqnarray}
\Omega_{red} =  m d\xi \wedge dq,
\end{eqnarray}
where
\begin{eqnarray}
\xi = \int^I \frac{dx}{\sqrt{1+x^2}} = \log (I + \sqrt{1+I^2}).
\end{eqnarray}
The coordinates $\xi $ and $q$ are global on $\Gamma_{red}$.
Moreover, they define a set of Cartesian coordinates for the
metric (\ref{metred}), since
\begin{eqnarray}
ds^2_{red} = \frac{1}{2} d \xi^2 + \frac{1}{2 m^2} dq^2
\end{eqnarray}
The metric and symplectic form on $\Gamma_{red} = {\bf R}^2$
correspond to that of the standard Gaussian coherent states. The
path-integral then leads to the overlap kernel
\begin{eqnarray}
\langle \xi, q| \xi' q' \rangle = \exp \left( im \xi'q - i m \xi
q' - \frac{1}{2} (\xi -\xi')^2 - \frac{m^2}{2} (q-q')^2 \right)
\label{rpkernel}
\end{eqnarray}
Even though the definition of the parameters $\xi$ and $q$
involved the choice of a coordinate system, the coherent states
constructed from the kernel (\ref{rpkernel}) are not. The reason
is that $q$ has a covariant definition as $q = \epsilon_{\mu
\nu}I^{\mu} X^{\nu}$, in terms of the alternating tensor
$\epsilon_{\mu \nu}$ in two dimensions, while $\xi \rightarrow \xi
+ c $ under a Lorentz boost in two dimensions, leaving the kernel
(\ref{rpkernel}) invariant up to a change of phase.

We may pull-back the kernel $\langle \xi, q| \xi' q' \rangle$ on
the constraint surface and write it in terms of $x = X^1$, $t =
X^0$ and $p = m I^1$, which are the natural variables in the
canonical description of time evolution. We obtain
\begin{eqnarray}
\langle x, p, t| x', p', t' \rangle = \exp \left( \frac{i}{m} p'
(\sqrt{m^2 + p^2} x - pt) - \frac{i}{m} p (\sqrt{m^2 + p'^2} x' -
p' t') \right.\nonumber \\ \left. - \frac{1}{2} [\log\frac{p +
\sqrt{m^2 + p^2}}{p' + \sqrt{m^2 + p'^2}}]^2 - \frac{1}{2}
(\sqrt{m^2 + p^2} x - p t - \sqrt{m^2 + p'^2}x' + p' t')^2
\right). \label{ckernel}
\end{eqnarray}
In the non-relativistic limit $p << m$, the overlap kernel reduces
to
\begin{eqnarray}
\langle x, p, t| x', p', t' \rangle = \exp \left( i p' (x - p/mt)
- i p(x' - p'/m t') - \nonumber \right. \\ \left. \frac{1}{2 m^2}
(p - p')^2 - \frac{m^2}{2} (x - p/ m t - x' + p'/m t')^2 \right)
\label{nonrel}
\end{eqnarray}
For $t = t'= 0 $ this kernel defines a Gaussian family of coherent
states of the Weyl group, typical in the description of
non-relativistic particles. However, for $t \neq t'$. $\langle x,
p, t| x', p', t' \rangle $ does not correspond to matrix elements
$\langle x, p|e^{-i\hat{H}(t - t')} | x' p' \rangle$ of any
self-adjoint Hamiltonian $\hat{H}$.

The coherent states (\ref{rpkernel}) are {\em instantaneous}: they
are properly defined only on the reduced state space, in which the
Hamiltonian vanishes due to constraints. For this reason the
time-dependence in the arguments of (\ref{ckernel}) reflects
solely the relation of the parameters on the reduced state space
to a coordinate that is usually considered to play the role of
time--see \cite{Sav99, SavAn00} for an interpretation. The
coherent states are covariant under the Poincar\'e group; for any
element $g$ of the Poincar\'e group, one may define the unitary
operator $\hat{U}(g)$ as
\begin{eqnarray}
\hat{U}(g) |z \rangle = | g \cdot z \rangle,
\end{eqnarray}
where $z \in \Gamma_{red}$ and $g\cdot z$ denotes the symplectic
action of the Poincar\'e group on $\Gamma_{red}$. For the one
parameter subgroup of time-translations  (the $0$-direction) in
particular, we obtain the transformation
\begin{eqnarray}
|\xi, q \rangle  \rightarrow |\xi, q + \sin \xi s \rangle,
\end{eqnarray}
which can be easily checked to be unitary. The matrix elements of
the Hamiltonian in the coherent state basis have no relation to
the pull-back of the coherent state overlap kernel on $C$, Eq.
(\ref{ckernel}). The parameter $t = X^0$ cannot be identified with
the time parameter of Schr\"odinger's equation, a fact that is
also responsible for the non-definability of a covariant position
operator for relativistic particles \cite{AnSav03}.

\paragraph{Minisuperspace models} Minisuperspace models are quantum
cosmological models characterised by a canonical state space ${\bf
R}^{2n}$ with configuration variables $q^a$ and conjugate momenta
$p_a$ and a by constraints of the form
\begin{eqnarray}
\frac{1}{2} g_{ab}(q) p^a p^b + V(q) = 0,
\end{eqnarray}
in terms of a metric $g$ on the configuration space with signature
$-++ \ldots +$. In many aspects their structure is similar to that
of the relativistic particle. There exist, however, no spacetime
covariance argument to predetermine a form of the metric. We have
applied the present method of quantisation to minisuperspace
models in Ref. \cite{AnSav04}, where we studied in detail  a
Robertson-Walker universe with a scalar field.

\subsection{Divergent points and orbifold structure}
In many systems with first-class constraints, there may exist
exceptional points of the reduced state space that correspond to
orbits of reduced dimensionality than the generic orbit. This is
usually the case, when specific subsets of the constraint surface
are invariant under some (or even all) the first-class
constraints. The symplectic structure on such points is typically
divergent, with the result that the reduced state space is not
well defined as a classical symplectic manifold. This is, in fact,
one of the reasons that Dirac quantisation is preferred over
RSSQ--one does not have to deal with such singular orbits in the
Dirac method.

We argue here that such divergent points do not pause any problem
in the quantisation scheme through coherent states that is
proposed here. The orbits corresponding to singular points in the
reduced state space are absorbed in a redefinition of the reduced
state space that is imposed by the consideration of the Riemannian
structure.

We shall consider for simplicity the case of a system with a
single first-class constraint. Each orbit of the constraint
surface $C$ is characterised by coordinates $\zeta^i$ that are
constant along the orbit--and thus project on the reduced state
space-- and a single gauge coordinate $\lambda$ distinguishing
points along the orbit. Let us assume that there exist a specific
point on $C$, determined by the coordinates
$(\zeta^i_0,\lambda_0)$, that is left invariant under the action
of the  constraint. The corresponding orbit $\gamma_0$ consists
then only of the single point $(\zeta^i_0, \lambda_0)$, while a
generic orbit is characterised by varying $\lambda$, namely it
corresponds to a line on $C$. There exist other orbits $\gamma'_0$
characterised by the same parameters $\zeta^i_0$ with $\gamma_0$:
they correspond to values of $\lambda > \lambda_0$ and $\lambda <
\lambda_0$. There will be two orbits if $\lambda$ runs to the full
real axis and only one orbit if $\lambda$ is a periodic variable.

To compute the metric on the reduced state space, one has to
calculate the distance function between different orbits. The
determination of the distance between $\gamma_0$ and and the other
orbits $\gamma_0'$ of the same value of $\zeta^i$, is the infimum
of the distance of the points of $\gamma_0'$ from the point
$(\zeta^i, \lambda_0)$. The infimum will be always achieved for
$\lambda \rightarrow \lambda_0$, implying that the distance
between $\gamma_0$ and $\gamma_0'$ will be equal to zero.
 Consequently, the Wiener process constructed from
the metric on $\Gamma_{red}$ will fail to distinguish between
$\gamma_0$ and $\gamma'_0$. The path-integration will, therefore,
treat all orbits with the same value of $\zeta^i$ as a single
point. The coherent states, therefore, will only be parameterised
by the regular points of $\Gamma_{red}$. Recalling the relation
between the state space metric and the uncertainty relation, one
may say that the quantum uncertainties essentially wash out any
divergences related to {\em accidental} symmetries of the
classical constraint surface.

The argument we provided here is very general and immediately
generalises to systems with more than one constraint functions.
For a detailed example (a Robertson-Walker minisuperspace model)
 the reader is referred to
\cite{AnSav04}.

\section{Conclusions}

We have presented here a variation of the reduced-state-space
quantisation procedure that is based on the
coherent-state-path-integral quantisation, developed by Klauder.
The key point of our construction is that a metric $ds^2_{red}$ on
the reduced state space may be constructed purely geometrically
from a metric on the unconstrained state space $\Gamma$. The
metric $ds^2_{red}$ may then be employed for the definition of a
path integral for the degrees of freedom in the reduced state
space.

The advantages of this method are the following:

\vspace{0.3cm}

i. While the method incorporates the constraints at the classical
level, it can be easily related to Dirac quantisation. The
identification of a metric on the reduced state space is
constructed uniquely from a metric on $\Gamma$ (or on the
constraint surface $C$), serves to define a map from a kinematical
Hilbert space (related to the degrees of freedom on $\Gamma$) to a
physical Hilbert space that corresponds to the true degrees of
freedom. One may therefor enjoy all benefits of Dirac
quantisation, for example the existence of quantum operators for
kinematical variables, without any of its drawbacks.

\vspace{0.25cm}

 ii. The singular points of the reduced state space
are 'smeared out' in the quantum theory, and do not appear as
parameters in the coherent states.

\vspace{0.25 cm}

iii. Finally, the method is purely geometrical; the basic
ingredient is the identification of the distance between
constraint orbits. As such it is particularly suitable for
theories, in which the state space involves geometric or
combinatorial objects: for example, approaches to canonical
quantum gravity that involve discretisation.

\section*{ Acknowledgements} This work was supported by a Marie
Curie
 Reintegration Grant of the European Commission and a research
grant from the Empirikion Foundation. I would also like to thank
N. Savvidou for many important discussions and remarks.


\begin{thebibliography}{}

\bibitem{KlDa84}  J. R. Klauder and I. Daubechies, {\em Quantum
Mechanical Path Integrals with Wiener Measures for all Polynomial
Hamiltonians}, Phys. Rev. Lett. 52, 1161 (1984);  I. Daubechies
and J. R. Klauder, {\em Quantum Mechanical Path Integrals with
Wiener Measures for all polynomial Hamiltonians: 2}, J. Math.
Phys. 26, 2239, (1985).

 \bibitem{Kla88} J. R. Klauder, {\em Quantization is Geometry,
 After All}, Ann. Phys. 188, 120, 1988; {\em Geometric Quantization
  from a Coherent State Viewpoint}, quant-ph/9510008; S. V.
  Shabanov and J. R. Klauder, {\em Path Integral Quantization and Riemannian-Symplectic
  Manifolds},  Phys.  Lett. B435, 343, (1998).

  \bibitem{RoSm95} C. Rovelli and L. Smolin, {\em Discreteness of
  Area and Volume in Quantum Gravity}, Nucl.Phys.B442, 593,
  (1995).

  \bibitem{Kla97} J. R. Klauder, {\em Coherent State Quantization of
Constraint Systems}, Ann. Phys. 254, 419,  (1997).

\bibitem{Klaother} J. R. Klauder, {\em Universal procedure for Enforcing Quantum
Constraints}, Nucl.Phys. B547,397, (1999);  A. Kempf and J. R.
Klauder, {\em On the Implementation of Constraints through
Projection Operators}, Phys. A34, 1019 (2001).

\bibitem{Ish84} C. Isham, {\em Topological and Global Aspects of
Quantum Theory},  Les Houches Rel.School 1983, 1059.

\bibitem{AnSav03} C. Anastopoulos and N. Savvidou,{\em The Role of
Phase Space Geometry in Heisenberg's Uncertainty Relations},
 Ann. Phys. 308, 329, (2003).


\bibitem{Kl81} J. R. Klauder, {\em Constructing Measures for Spin-variable Path
Integrals}, J. Math. Phys.23, 1797, (1982).

\bibitem{Sav99} K. Savvidou, {\em The action operator in
continuous time histories}, J. Math. Phys.  40, 5657, (1999);  K.
Savvidou;
\newblock  {\em Continuous Time in Consistent Histories},
\newblock  {PhD Thesis} in gr-qc/9912076, (1999).

\bibitem{SavAn00} K. Savvidou and C. Anastopoulos, {\em  Histories
quantisation of parametrised systems: I. Development of a general
algorithm}, Class.\ Quant.\ Grav. 17, 2463, (2000).



\bibitem{AnSav04} C. Anastopoulos and N. Savvidou,  {\em Minisuperspace Models in Histories
Theory}, gr-qc/0410131.




\end{thebibliography}
\end{document}